\documentclass{epl}
\usepackage{amsmath} 
\usepackage{amsfonts}
\usepackage{amssymb}

\title{Periodically driven stochastic un- and refolding transitions of biopolymers}
\shorttitle{Periodic transitions of biopolymers}
\author{O. Braun\inst{1} \and U. Seifert\inst{1}} 
\institute{ \inst{1}{II. Institut f\"ur Theoretische Physik, Universit\"at 
Stuttgart, 70550 Stuttgart, Germany}}

\newcommand{\eqa}{\begin{eqnarray}}
\newcommand{\eqe}{\end{eqnarray}}
\newcommand{\la}{\langle}
\newcommand{\ra}{\rangle}
\newcommand{\avr}[1]{\la #1 \ra}
\newcommand{\dt}{{\frac{\rm d}{{\rm d}t}}}

\newcommand{\A}{\alpha}
\newcommand{\B}{\beta}

\newcommand{\D}{\delta}
\newcommand{\DD}{\Delta}

\newcommand{\W}{\omega}
\newcommand{\T}{\tau}

\newcommand{\Si}{\sigma}

\newcommand{\p}{\Phi}
\newcommand{\K}{\kappa}

 {Biomolecules: structure and physical properties}
 {Scanning tunneling and atomic force microscopy}

\begin{document}

\maketitle

\begin{abstract} 
Mechanical single molecule experiments probe the energy profile of biomolecules.
We show that in the case of a profile with two minima (like
folded/unfolded) periodic driving leads to a stochastic resonance-like
phenomenon. We demonstrate that the analysis of such data can be used
to extract four basic parameters of such a transition and discuss
the statistical requirements of the data acquisition. As advantages of the
proposed scheme, a polymeric linker is explicitly included and thermal
fluctuations within each well need not to be resolved. 
\end{abstract}

Force spectroscopy has been used extensively to probe mechanically the 
interactions within a single
biomolecule, like RNA and DNA, peptides or proteins by applying an
externally controlled load on such a structure; for reviews see
\cite{Mehta99, Clausen-Schaumann00, Evans01, Merkel01, Rief02, Strick03, Bustamante03}.  Since in these 
experiments typically the extension along one coordinate $r$ is controlled, 
the data and results  are rationalized in terms of the notion of a  
(free) energy
profile $G(r)$ along this coordinate. A major challenge for future
improvements is to design these experiment is such a way that as much 
information as possible on such a profile can be extracted reliably from 
the data.

\begin{figure}
\twofigures[scale=0.5]{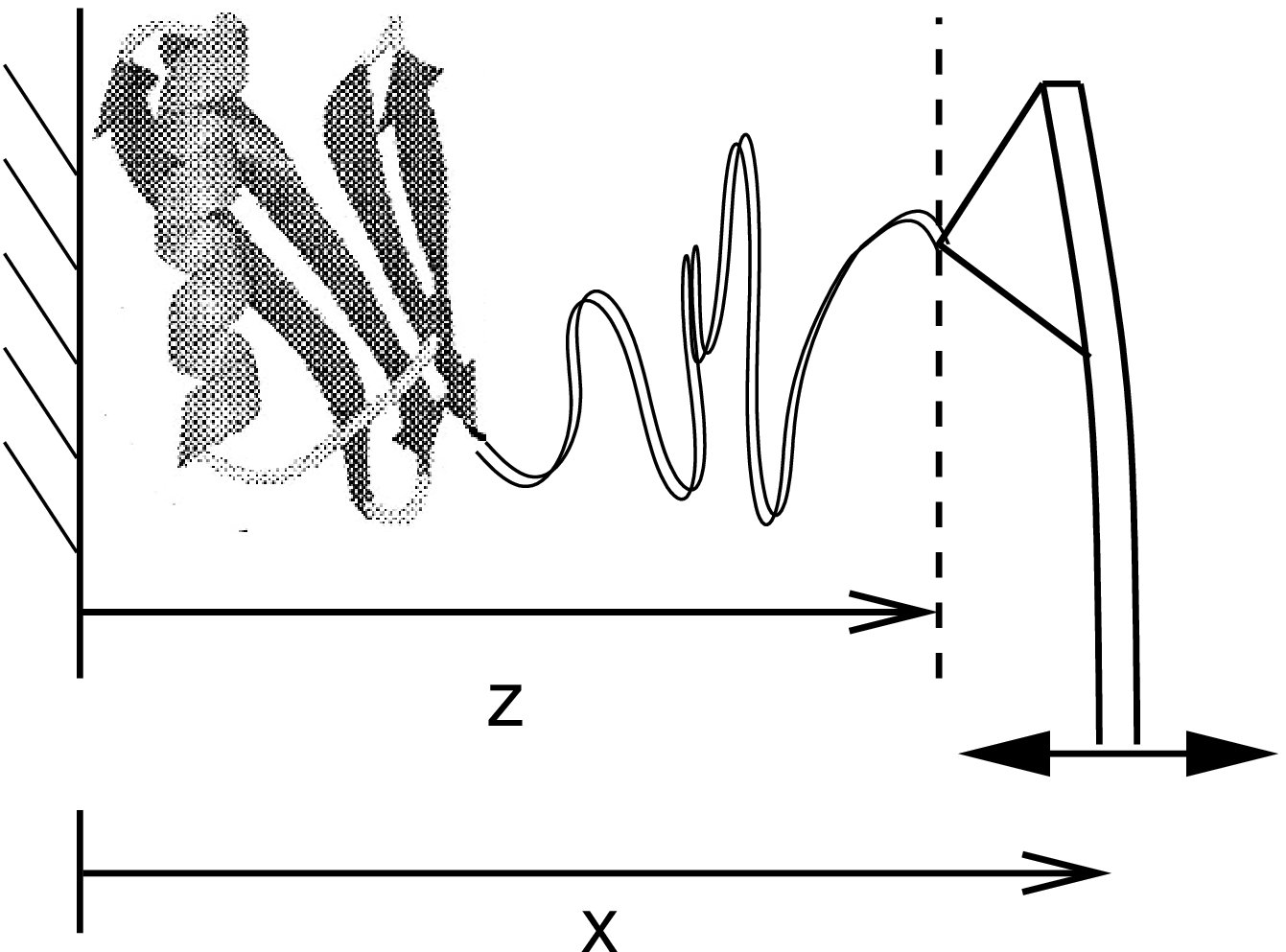}{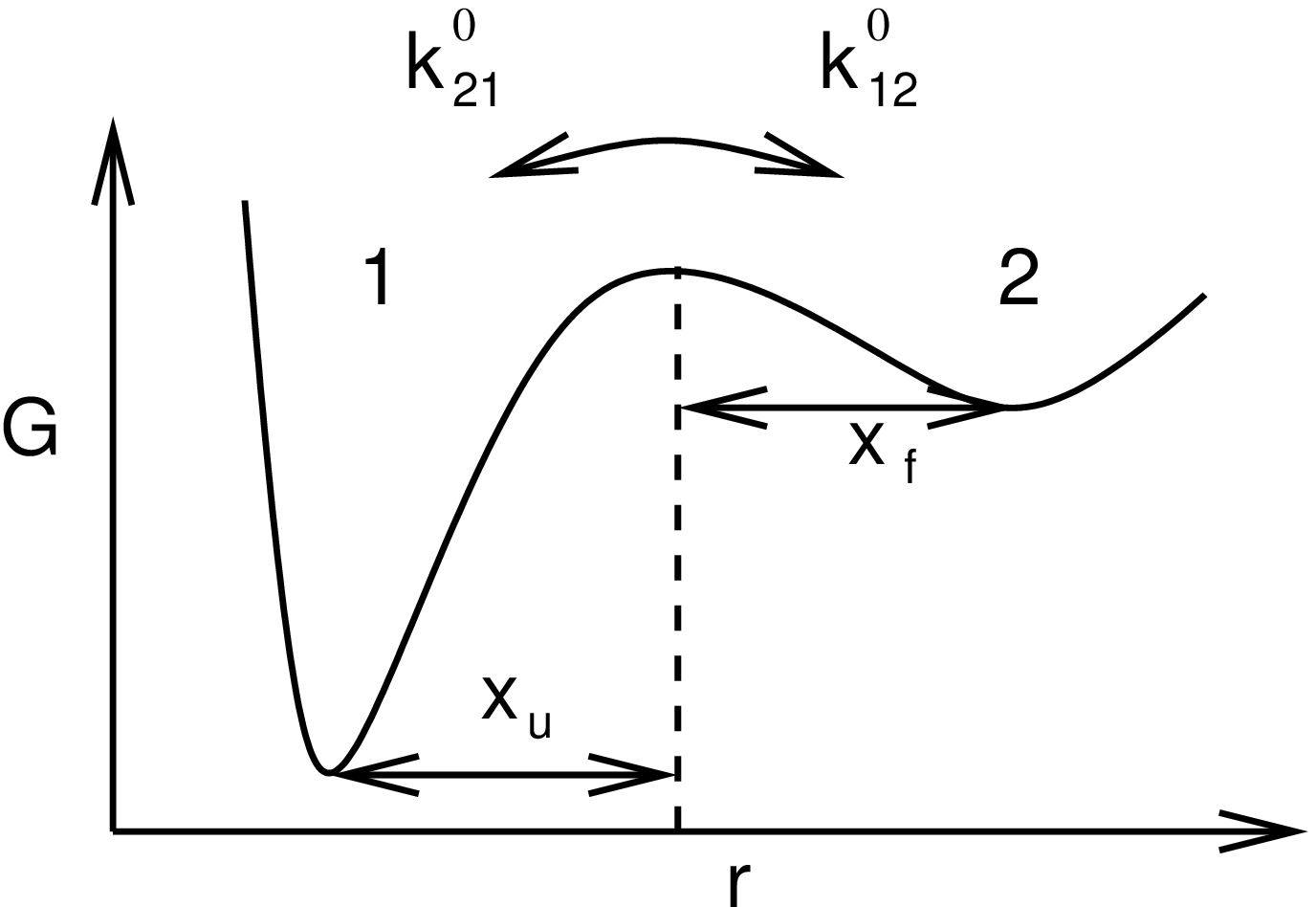}
\caption{Sketch of an atomic force spectroscopy experiment. The protein
(thick lines) is connected via a polymeric linker
(thin line) to the
 tip of an AFM cantilever at  $z(t)$ whose base at $x(t)$ is externally 
controlled.}
\label{aufbau.fig}
\caption{Schematic view of the underlying 
 free energy potential $G(r)$ as a function of the reaction
coordinate $r$.  The first minimum 
represents the folded state, whereas the second shallow minimum 
represents the unfolded state of the biopolymer. $x_{\rm u}$ and $ x_{\rm f}$ denote the
respective distances from the minima to the barrier and $k_{ij}^0$ the 
spontaneous transition rates from state $i$ to $j$.}
\label{bez1b.fig}
\end{figure}

 In the paradigmatic situation
of an experiment using an AFM cantilever, the biomolecule is attached via 
a polymeric linker to the tip of the AFM whose base is driven according
to a certain protocol, see fig.\;\ref{aufbau.fig}.
Up to now, in most applications in force spectroscopy, the force is applied
according to a linear ramp protocol $x(t)=x_0+vt$. Here, $x_0$ is the 
cantilever position at time $t=0$ and $v$ is the constant ramp velocity.
From the peak of the distribution of unfolding forces, one can extract two
parameters of the underlying free energy landscape: the distance $x_{\rm u}$ 
from the first minimum to the maximum 
and the spontaneous off-rate $k_{12}^0$ \cite{Evans97}, 
see fig.\;\ref{bez1b.fig}. These quantities characterize mainly the 
unfolding part of the profile. If one wants more information about 
the potential $G(r)$,
in particular on quantities determining the refolding events,
one has to resort to a more sophisticated analysis.

 In the latest development,
analyzing the distribution of the applied work spent in the unfolding 
process yields the full underlying profile using Jarzynski's relation 
\cite{Jarzynski97, Hummer01, Liphardt02}. In this context, we have shown that  a periodic rather 
than a linear ramp improves the reconstruction 
of the profile significantly \cite{obahus}. Such an analysis based on 
Jarzynski's relation,
however, requires first that the position of the tip $z(t)$ is measured very 
accurately. Second, the reconstructed energy profile refers to the 
coordinate $z$ which includes the linker molecule. It is nontrivial to
extract from such a profile $G(z)$ the quantity $G(r)$ which is of main 
interest.

In this letter, we propose a method on an intermediate level of
sophistication, which yields, apart from  $x_{\rm u}$ and $ k_{12}^0$,  the 
corresponding two 
quantities $ x_{\rm f}$ and $ k_{21}^0$ of the refolding transition.
We employ a periodic ramp which leads to a stochastic resonance-like
phenomenon \cite{Gammaitoni98}. As advantages of our new scheme, we include 
the linker and do not require to analyze the full stochastic trajectory
$z(t)$ as it is necessary in approaches based on Jarzynski's relation. 
It is sufficient to observe the sequence of folding and unfolding 
events reflected in this coordinate as explained below.
Our proposal extends a previous approach where spontaneous transitions under
constant force are used to extract these parameters \cite{Liphardt01}. 
The main advantage
of our dynamical approach is that stochastic resonance leads to an
enhancement of the transition. Moreover more control parameters such as
amplitude and frequency of the driving are available to optimize the 
data acquisition. Finally no feed-back loop in order 
to impose a constant force is necessary.

We first recall that the spontaneous transition rates $k_{12}^0$ and 
$k_{21}^0$ for unfolding and refolding are modified 
by a force $F$  according to \cite{Bell78, Evans97, Rief98b}
\eqa
\label{1.1}
k_{12}&=&k_{12}^0 \exp [\B x_{\rm u}F(z,L)]\\ \label{1.2}
k_{21}&=&k_{21}^0 \exp [-\B x_{\rm f}F(z,L)]\, \, 
\eqe 
where $\B$ is the inverse temperature. The force   $F(z,L)$ is transmitted
from the cantilever to the biomolecule
by a polymeric linker which we model by a worm-like chain
as  \cite{Marko95}
\eqa \label{wlc}
F(z,L)=\frac{1}{\B L_{\rm p}}\left(\frac{1}{4(1-z/L)^2}-\frac{1}{4}+\frac{z}{L}\right)\, \, ,
\eqe
 where  $L_{\rm p}$ is the persistence length. The contour length $L$ 
depends on the configuration. In the folded state, it is $L=L_0$, in the 
unfolded state $L=L_0+\DD L$, where $\DD L$ is the length gain of the contour 
length upon unfolding. 

In each state, the force $F_{\rm c}$ measured by the cantilever with harmonic spring 
constant $k_{\rm c}$ has to balance the polymeric force
\eqa
\label{force}
F_{\rm c}\equiv k_{\rm c}(x-z)= F(z,L) \, \, .
\eqe
Thus, the $z$-coordinate can be calculated for given externally controlled 
cantilever base position $x$. In general eq. (\ref{force}) can not be solved 
analytically for $z(x)$. 
In force spectroscopy on biopolymers, however, the linker is almost stretched 
out for a typical unfolding force of about $50$-$150$ pN.
Eqs. (\ref{wlc}) and (\ref{force}) then reduce to the cubic equation
\eqa
\label{force2}
k_{\rm c} (x-L+\D)=\frac{L^2}{4\B L_{\rm p} \D^2}\, \, ,
\eqe
for $\D\equiv L-z$, which denotes the length stored in the fluctuations.
 Hence the functions $z=z(x,L)$ or 
$\D=\D(x,L)$ also depend on the current length of the polymer.

We now apply a periodic protocol of the cantilever position $x$ according 
to  
\eqa
x(t)=x_0+x_{\rm{a}} \cos(\W t)\, \, ,
\eqe
with the offset extension  $x_0$, the amplitude $x_{\rm{a}}$ and frequency $\W$
\cite{obahus}. 
In general, the  
periodicity of $x(t)$ leads to an anharmonic periodicity of $\D(t)$ via eq. 
(\ref{force2}), but the leading term for $x_{\rm{a}} \ll x_0$
 is given by the first harmonics
as 
\eqa
\D(t)\approx \D_0+\D_{\rm{a}} \cos(\W t)\, \, ,
\eqe
where $\D_0$ is the solution of eq. (\ref{force2}) with $x=x_0$. The amplitude 
$\D_{\rm{a}}$ is given by
\eqa
\D_{\rm{a}}=-\frac{x_{\rm{a}}}{3+2(x_0-L)/\D_0}\, \, .
\eqe
To first order in $\D_{\rm{a}}/\D_0$, the force acting on the biomolecule
exhibits the same periodicity
\eqa 
F(\D,L)\approx\frac{L^2}{4\B L_{\rm p} \D_0^2}\left(1-\frac{2\D_{\rm{a}}}{\D_0}\cos(\W t)\right)\equiv F_0+F_{\rm{a}}\cos(\W t).
\label{us:force}
\eqe
For the functions $F_0$ and $F_{\rm{a}}$ we will have to distinguish between 
the folded the unfolded configuration.

The probability $\p(t)$ that the molecule is in the folded state
obeys the master equation
\eqa
\label{master1}
\dt{\p}=-(k_{12}+k_{21})\p+k_{21},
\eqe
where the rates depend on the force (\ref{us:force}) through (\ref{1.1}) and 
(\ref{1.2}).
For small loading, an analytical solution can be calculated 
in linear order in $x_{\rm{a}}/x_0$ by splitting the solution of 
the master equation (\ref{master1}) into a time-independent part $\p_0$ and a 
time-dependent part $\Pi(t)$ as
\eqa
\label{solution}
\p(t)=\p_0+\Pi(t)\, \, .
\eqe
The stationary part is given by
\eqa
\label{res0}
\p_0=\frac{\K_{21}}{\K_{12}+\K_{21}}\, \, , 
\eqe
with 
\eqa
\K_{12}\equiv k_{12}^0\exp [\B x_{\rm u} F_0(\D_0,L_0)] \qquad \mbox{and}\qquad
\K_{21}\equiv k_{21}^0 \exp [-\B x_{\rm f} F_0(\D_0,L_0+\DD L)]\, \, ,
\eqe
which are the time-independent
contributions to the transition rates due to the preloading. 
On the intrinsic time scale $1/\Lambda\equiv (\K_{12}+\K_{21})^{-1}$, 
the time dependent part of  $\p(t)$ approaches a limit cycle
\eqa
\label{res3}
\Pi(t)\equiv \Pi_0 \cos (\W t-\lambda)=
\frac{\A}{\sqrt{\W^2+\Lambda^2}} \cos (\W t -\lambda)\, \, ,
\eqe
 with 
\eqa
\A\equiv -(\B x_{\rm u} F_{\rm{a}}(L_0)+ \B x_{\rm f} F_{\rm{a}}(L_0+\DD L))\K_{12}\p_0
\eqe
and the phase shift 
\eqa
\label{phase}
\lambda\equiv \arctan (\W/\Lambda)\, \, .
\eqe 
 The solution of eq. (\ref{master1}) is thus a 
periodic response of the probability of occupation $\p$ of the folded state.

For fixed $x_0$ and fixed $\W$ the solution (\ref{solution}) yields three 
parameters $\A, \Lambda$ and $\Phi_0$ for the four unknowns $x_{\rm u}, x_{\rm f}, k_{12}^0, k_{21}^0$. For a determination of these crucial parameters,
 more data are needed, which can easily be obtained by varying 
the offset extension $x_0$ and the frequency $\W$. Such a variation leads 
to an interesting resonance phenomenon.

\begin{figure}
\twofigures[scale=0.55]{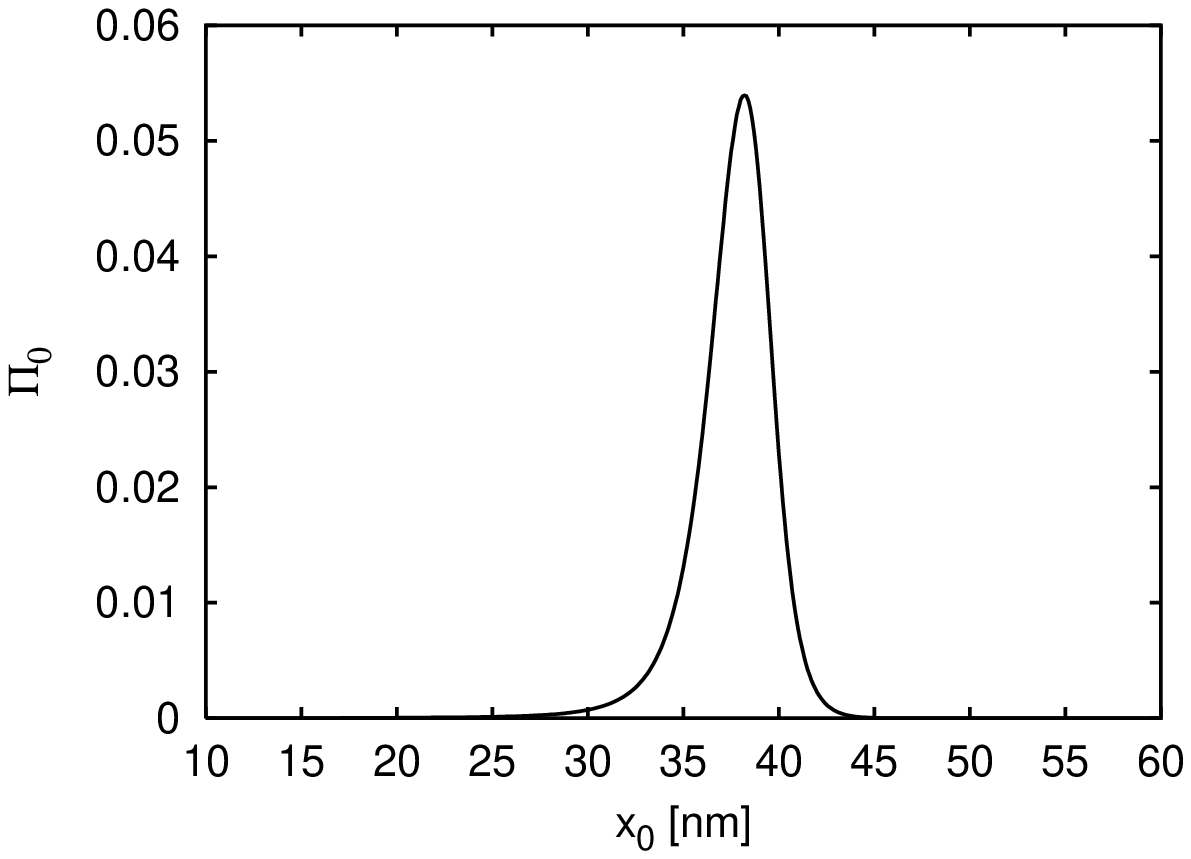}{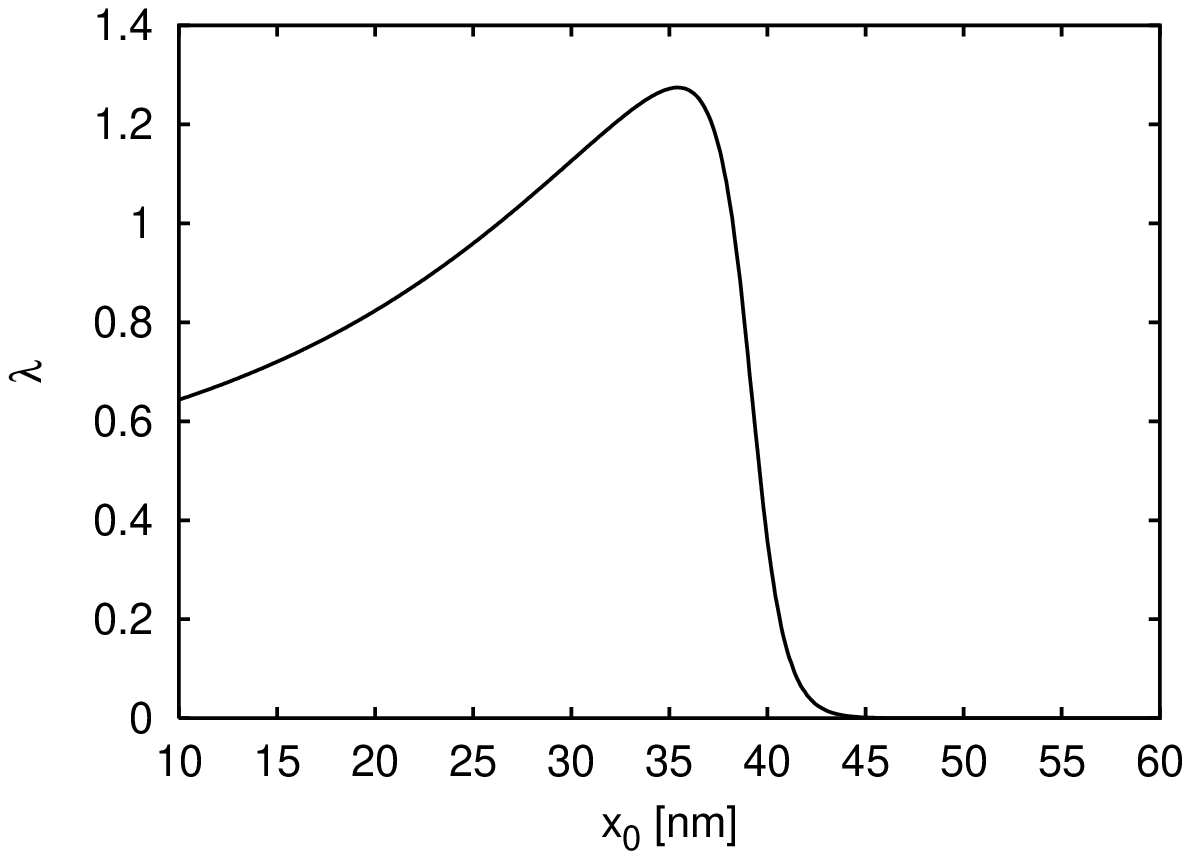}
\caption{The analytical solution of the response amplitude of the occupation probability of the folded state $\Pi_0$ as a function the offset extension $x_0$. The parameters are $k_{21}^0=3 \un{s^{-1}}$, $k_{12}^0=4.6\cdot 10^{-4} \un{s^{-1}}$, $x_{\rm u}=0.3 \un{nm}$, 
$x_{\rm f}=1 \un{nm}$, $L_{\rm p}=0.42\un{nm}$ , $L_0=40\un{nm}$ , $\DD L=28.5\un{nm}$, $\omega=1 \un{s^{-1}}$, $x_{\rm{a}}=0.5\un{nm}$, $k_{\rm c}=5 \un{pN/nm}$, taken from \cite{oberhauser98} for a tenascin FN-III segment.}
\label{paper1a.fig} 
\caption{The phase shift $\lambda$ of the occupation probability over the offset extension $x_0$ for the same parameters as in fig.\;\ref{paper1c.fig}.}
\label{paper1c.fig} 
\end{figure}

It is one of the well known features of  periodically driven stochastic 
transitions, that in contrast to a deterministic mechanical system,
 the response amplitude  depends on the frequency in a non-resonant 
manner \cite{Gammaitoni98}.  
For low frequencies, where  the stochastic transition can follow the externally
driven loading, the amplitude is larger than for 
large frequencies, see eq. (\ref{res3}).
Both, the response amplitude $\Pi_0$ of the periodic part of
the occupation probability and the phase shift $\lambda$ 
show a clear maximum as a function of the offset $x_0$, 
see figs.\;\ref{paper1a.fig} and \ref{paper1c.fig}. 
The maximal phase shift $\lambda^*$ follows from the condition
$\K_{12}' +\K_{12}'=0$, where the prime denotes the derivative with respect 
to $x_0$. This yields the ratio of the geometric parameters $x_{\rm f}$ and $x_{\rm u}$ 
as 
\eqa
\label{maximum}
\frac{x_{\rm f}}{x_{\rm u}}=\frac{\K_{12} {F_0'(L_0)}}{\K_{21} {F_0'(L_0+\DD L)}} \, \, ,
\eqe
where $F_0'(L_0)$ and $F_0'(L_0+\DD L)$ are the derivatives of the force with 
respect to $x_0$
with folded contour length and unfolded contour length respectively.
With eq. (\ref{force2}), we get explicitly
\eqa
F_0'(L)&=&\frac{L^2}{2 \B L_{\rm p} {\D_0^*}^2(\D_0^*+2(\D_0^*+x_0^*-L))}\, \, ,
\eqe
where $x_0^*$ is the position of the maximum of the phase shift and 
$\D_0^*=\D_0(x_0^*,L)$.
Eq. (\ref{maximum}) then  yields the ratio $x_{\rm u}/x_{\rm f}$
of the underlying free energy profile as a fourth relation which can be used
to determine $x_{\rm u}, x_{\rm f}, k_{12}^0, k_{21}^0$. Of course, many other fitting
schemes can be envisaged to extract these crucial parameters.

Up to now, we have not yet addressed explicitly
 the question of how to obtain the 
probability $\Phi(t) $ from real measurements and how these
 experiments have to
be designed to deliver the necessary data most efficiently.
In principle, the occupation probability $\Phi(t)$ has to be 
measured via an ensemble average of trajectories. The ensemble average,
however, can be replaced by a phase sensitive time average of a 
single trajectory, because $\p(t)$ is periodic with period time $T=2\pi/\W$ 
after a relaxation time for the offset expansion of the order 
of $t\sim 1/\Lambda$. Comparing these two averaging schemes, the advantage 
of the phase sensitive time average is obviously that the relaxation time 
has to be accounted for only once.

According to our assumptions, the trajectories $z(t)$ are 
bistable, so the conformation 
(folded or unfolded) of the biopolymer can be 
determined unambiguously from $z(t)$ since each transition leads to
a jump in $z(t)$ which is larger than the thermal fluctuations within
each minimum. We assign a weight $p(t)=1$ if the molecule at time
$t$ is in  the folded state and
$p(t)=0$ if it is in the unfolded state. The probability of occupation 
$\p(t)$ of the folded state can then be estimated by
\eqa
\label{measure}
\p_N(t)=\frac{\sum_{n=0}^{N-1} p_n(t)}{N}\, \, ,
\eqe
where $N=T_{\rm tot}/T$ is the number of periods with the total measuring 
time $T_{\rm tot}$ and $p_n(t)\equiv p(t+nT)$ and $0<t<T$.
We now optimize $\W$ and $T_{\rm tot}$ for the least statistical error.

\begin{figure}
\onefigure[scale=0.55]{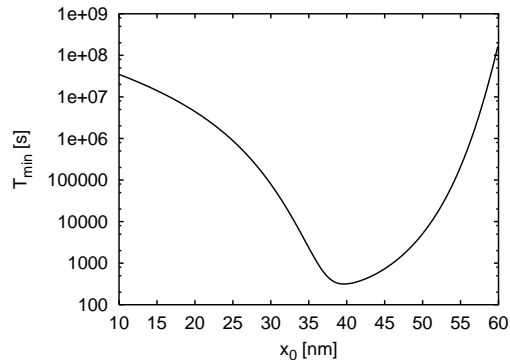}
\caption{The minimum total measuring time $T_{\rm min}$ over the offset extension $x_0$ for the optimized frequency $\W^*\simeq\Lambda$ and the same parameters as
in figs.\;\ref{paper1a.fig} and \ref{paper1c.fig}.}
\label{paper1b.fig} 
\end{figure}

The autocorrelation of $p(t)$  up to first order 
in $x_{\rm{a}}/x_0$ is given by \cite{Gammaitoni98}
\eqa
\label{kovarianz}
\mbox{cov}(p(t),p(t+\T))\equiv 
\avr{(\avr{p(t)}-p(t))(\avr{p(t+\T)}-p(t+\T))}\approx
\sigma_{\rm p}^2 e^{-\Lambda \T}\,\, ,
\eqe
with the single measurement error of a two state system $\Si_{\rm p}^2(t)=
 \p(t)(1-\p(t))$. 
We consider the $N$-dependent stochastic variable $\p_N(t)$, which fluctuates 
around the mean average value $\p(t)$ with a 
variance $\Si_{\p_N}^2=\sum_{n=0}^{N-1} \Si_{p}^2 (\partial \p_N/
\partial p_n)^2+\sum_{n\ne m}^{N-1} (\partial \p_N/\partial p_n)
(\partial \p_N/\partial p_m) \mbox{cov}(p_n,p_m)$, where we omitted 
the $t$-dependence. Together with eq. (\ref{kovarianz}) we get approximately 
a geometric sum, which yields
\eqa
\Si_{\p_N}^2(t)\simeq \frac{\Si_{\rm p}^2(t)}{N(1-e^{-\Lambda T})}\, \, .
\eqe
The square root of the variance of this stochastic variable can now be 
identified with the measuring error, which should be smaller than the response 
amplitude $\Pi_0$. Thus, we get the condition 
$\sqrt{\Si_{\p_N}^2}\lesssim |\Pi_0|$. The total measuring time must therefore obey
\eqa
\label{totaltime}
T_{\rm tot} \gtrsim \frac{2\pi \p_0(1-\p_0)(\W^2+\Lambda^2)}
{\A^2 \W (1-e^{-\Lambda 2 \pi/ \W})}\,\, ,
\eqe
where we have replaced the phase sensitive $\p(t)(1-\p(t))$ by $\p_0(1-\p_0)$,
which is permissible in zeroth order of ${\rm O}(x_{\rm{a}}/x_0)$.
The right hand side of equation (\ref{totaltime}) has a minimum at
$\W^*\simeq \Lambda$, if the exponential correction in the denominator
is neglected. 
The total measuring time for the optimized frequency $\W^*$ is
$T_{\rm min}\equiv T_{\rm tot}(\W^*)=4\pi \K_{12}\K_{21}/(\Lambda \A^2)$, which
strongly depends on the offset extension $x_0$, see 
fig.\;\ref{paper1b.fig}.
The minimum around $x_0^*$ is a signature 
of a stochastic resonance-like phenomenon. Usually in stochastic 
resonance, the potential and a periodic driving input is given while the
noise strength is controlled. One then finds an optimal noise strength
for the best amplification of the input signal. In our case the 
strength of the noise is fixed by the temperature, whereas the driving 
force can easily be manipulated by adjusting offset $x_0$, amplitude
$x_{\rm{a}}$ and frequency $\W$. Figs.\;\ref{paper1a.fig} and \ref{paper1b.fig}
show that where the response amplitude $\Pi_0$ 
as a function of offset $x_0$ is the largest the total measuring 
time $T_{\rm min}$ is the smallest. Hence, the signal to noise ratio is the 
best for this choice of offset and optimized frequency. Moreover, 
it should be noted that typically in 
stochastic resonance the bistable potential is symmetric whereas in our case 
it is intrinsically asymmetric due to the molecular interactions.

In summary, we have shown that the two spontaneous transition 
rates $k_{12}^0$, $k_{21}^0$
and the two most relevant length scales $x_{\rm u}$ and $x_{\rm f}$ of a biomolecular 
conformational transition can be extracted from data obtained by periodically
driving these transitions. Our analysis focuses on the measured occupation 
probability of one state, which exhibits a stochastic resonance-like 
phenomenon as a function of the control parameters of a typical 
AFM experiment. Compared to more sophisticated approaches based on
Jarzynski's relation, our scheme does not require 
to resolve thermal fluctuation within each minimum. Moreover, the 
linker molecule is explicitly included.
Future applications of our scheme will include more complex transitions and 
linear sequences of similar transitions in the un- and refolding of
multi-domain biopolymers. Most important, however, are experimental tests of
periodic loading which we hope to encourage with this study.
\acknowledgments

We thank A. Hanke for fruitful interactions.


\begin{thebibliography}{0}



\bibitem{Mehta99}
   \Name{Mehta A. D., Rief M., Spudich J. A., Smith D.A. \and Simmons R. M.}
   \REVIEW{Science}{283}{1999}{1689}.

\bibitem{Clausen-Schaumann00}
   \Name{Clausen-Schaumann H., Seitz M., Krautbauer R. \and Gaub H. E.}
   \REVIEW{Curr. Opin. Chem. Biol.}{4}{2000}{524}.
 
\bibitem{Evans01}
   \Name{Evans E.}
   \REVIEW{Ann. Rev. Biophys. Biomol. Struct.}{30}{2001}{105}.
 
\bibitem{Merkel01}
   \Name{Merkel R.}
   \REVIEW{Phys. Rep.}{346}{2001}{344}.
 
\bibitem{Rief02}
   \Name{Rief M. \and Grubm\"uller H.}  
   \REVIEW{ChemPhysChem}{3}{2002}{255}.

\bibitem{Strick03}
   \Name{Strick T. R., Dessinges M. N., Charvin G., Dekker N. H., Allemand J. F., Bensimon D. \and Croquette V.} 
   \REVIEW{Rep. Prog. Phys.}{66}{2003}{1}.

\bibitem{Bustamante03}
   \Name{Bustamante C., Bryant Z. \and Smith S. B.} 
   \REVIEW{Nature}{421}{2003}{423}.

\bibitem{Evans97}
   \Name{Evans E. \and Ritchie K.}
   \REVIEW{Biophys. J.}{72}{1997}{1541}.

\bibitem{Jarzynski97}
    \Name{Jarzynski C.}
    \REVIEW{Phys. Rev. Lett.}{78}{1997}{2690}.

\bibitem{Hummer01}
   \Name{Hummer G. \and Szabo A.}
   \REVIEW{Proc. Natl. Acad. Sci.}{98}{2001}{3658}.

\bibitem{Liphardt02}
  \Name{Liphardt J., Dumont S., Smith S. B., Tinoco Jr. I. \and Bustamante C.}
  \REVIEW{Science}{296}{2002}{1832}.

\bibitem{obahus}
   \Name{Braun O., Hanke A. \and Seifert U.}
   \REVIEW{cond-mat/0402496}{}{2004}{}.


\bibitem{Gammaitoni98}
   \Name{Gammaitoni L., H\"anggi  P., Jung P. \and Marchesoni F.}
   \REVIEW{Rev. Mod. Phys.}{70}{1998}{223}.

\bibitem{Liphardt01}
   \Name{Liphardt J., Onoa B., Smith S. B., Tinoco Jr. I. \and Bustamante C.}
   \REVIEW{Science}{292}{2001}{733}.
   
\bibitem{Bell78}
   \Name{Bell G. I.}
   \REVIEW{Science}{200}{1978}{618}.

\bibitem{Rief98b} 
   \Name{Rief M., Fernandez J. M. \and Gaub H. E.}
   \REVIEW{Phys. Rev. Lett.}{81}{1998}{4764}.



\bibitem{Marko95}
   \Name{Marko J. F. \and Siggia E. D.}
   \REVIEW{Macromolecules}{28}{1995}{8759}.

\bibitem{oberhauser98}
   \Name{Oberhauser A. F., Marszalek P. E., Erickson H. P. \and Fernandez J. M.}
   \REVIEW{Nature}{393}{1998}{181}.

\end{thebibliography}
\end{document}